\documentclass[final]{aipproc}
\layoutstyle{6x9}

\usepackage{amssymb,amsmath}
\def \eff{{\text{eff}}}

\begin{document}

%%%%%%%%%%%%%%%%%%%%%%%%%%%%%%%%%%%%%%%%%%%%%%%%%%%%%%%%%%%%%%%%%%%%%%%%%%%%%%%%%%%%%%%%
\title{Probing New Physics with   $\mathbf{b\to s \ell^+ \ell^-}$ and $\mathbf{ b\to s \nu \bar \nu}$ transitions}

\classification{13.20.He, 13.30.Ce, 12.60.-i}
\keywords      {Beyond Standard Model, B-Physics, Rare Decays}

\author{Michael Wick}{
address={Physik Department, Technische Universit\"at M\"unchen, D-85748 Garching, Germany}
}

%%%%%%%%%%%%%%%%%%%%%%%%%%%%%%%%%%%%%%%%%%%%%%%%%%%%%%%%%%%%%%%%%%%%%%%%%%%%%%%%%%%%%%%%
\begin{abstract}

The rare decay $B \to K^* (\to K \pi) \mu^+ \mu^-$ is regarded as one of the crucial channels for $B$ physics since its angular distribution gives access to many observables that offer new important tests of the Standard Model (SM) and its extensions. We point out a number of correlations among various observables which will allow a clear distinction between different New Physics (NP) scenarios. Furthermore, we discuss the decay $B \to K^* \nu \bar \nu $ which allows for a transparent study of $Z$ penguin effects in NP frameworks in the absence of dipole operator contributions and Higgs penguin contributions. We study all possible observables in  $B \to K^* \nu \bar \nu $ and the related $b \to s$ transitions $B \to K \nu \bar \nu $ and  $B \to X_s  \nu \bar \nu $ in the context of the SM and various NP models.
\end{abstract}
\maketitle
%%%%%%%%%%%%%%%%%%%%%%%%%%%%%%%%%%%%%%%%%%%%%%%%%%%%%%%%%%%%%%%%%%%%%%%%%%%%%%%%%%%%%%%%
\section{Introduction}
%%%%%%%%%%%%%%%%%%%%%%%%%%%%%%%%%%%%%%%%%%%%%%%%%%%%%%%%%%%%%%%%%%%%%%%%%%%%%%%%%%%%%%%%
The decays $B \to K^* (\to K \pi) \mu^+ \mu^-$ and $B \to K^* \nu \bar \nu$ are to some extent
complementary: combined with its charge conjugated counterpart, the all-charged final state of the first decay gives
access to a vast number of observables sensitive to CP violation. On the other hand, although the second decay yields only two observables,
it offers a unique possibility to a transparent study of $Z$ penguin effects. This is due to the massless, neutral final state, 
which goes along with an absence of non-perturbative contributions related to low energy QCD dynamics and photon exchanges.
What these decays have clearly in common is the important role they will play in the upcoming experiments such as LHCb and Super-B
facilities.

 \vspace{-0.8cm}
%%%%%%%%%%%%%%%%%%%%%%%%%%%%%%%%%%%%%%%%%%%%%%%%%%%%%%%%%%%%%%%%%%%%%%%%%%%%%%%%%%%%%%%%
\subsection{The $B \to K^* (\to K \pi) \mu^+ \mu^-$ decay}
%%%%%%%%%%%%%%%%%%%%%%%%%%%%%%%%%%%%%%%%%%%%%%%%%%%%%%%%%%%%%%%%%%%%%%%%%%%%%%%%%%%%%%%
%\subsubsection{Introduction}
A prediction for the observables of $B \to K^* (\to K \pi) \mu^+ \mu^-$ \cite{Altmannshofer:2008dz,Egede:2008uy,Bobeth:2008ij} involves mainly 
three theoretical ingredients: effective Hamiltonian, form factors and QCD factorization.
%\subsubsection{effective Hamiltonian}
While the effective Hamiltonian governing this decay is discussed in \cite{Altmannshofer:2008dz,Bobeth:2001jm}, we emphasize here that we also include scalar operators and lepton mass effects.
%\subsubsection{Form Factors}
The $B \to K^*$ matrix elements of the relevant operators can be expressed in terms of seven form factors depending
on the momentum transfer $q^2$ between the $B$ and the $K^*$ mesons. The well established technique of
QCD sum rules on the light cone (LCSR) that is applied here combines classic QCD sum rules with information on light cone distribution amplitudes in order to determine the form factors. The result is a set of form factors fulfilling all correlations required in the heavy quark limit.
%\subsubsection{QCD factorization}
In addition to terms proportional to form factors, the $B \to K^* \mu^+\mu^-$ amplitude
contains certain ``non-factorizable" contributions, which do not correspond to form factors. We include these QCD factorization corrections to NLO in $\alpha_s$ but LO in $1/m_b$. 
%\subsubsection{Observables}
The resulting angular distribution gives rise to twelve angular coefficient functions $I^{(a)}_i$ for both the original and the CP-conjugate mode. 
Instead of using these coefficients as fundamental observables, we use straight forward combinations of the $I^{(a)}_i$ and $\bar I^{(a)}_i$ to reduce the theoretical error
and separate CP-conserving and CP-violating NP effects. The twelve CP averaged
angular coefficients as well as the twelve CP asymmetries \cite{Altmannshofer:2008dz,Bobeth:2008ij} are given in terms of angular coefficient functions:
\begin{equation}
 S^{(a)}_i = \left( I^{(a)}_i + \bar I^{(a)}_i \right) \bigg/ \frac{d(\Gamma+\bar\Gamma)}{dq^2},~~~~~~~
 A^{(a)}_i = \left( I^{(a)}_i - \bar I^{(a)}_i \right) \bigg/ \frac{d(\Gamma+\bar\Gamma)}{dq^2}\,.
\end{equation}

Since this is a complete set of accessible observables, all previously considered observables, for example the forward-backward symmetry,
can be expressed straight forwardly in terms of new observables.
%\subsubsection{New Physics}
While in \cite{Altmannshofer:2008dz} several different models including the Littlest Higgs Model with T-Parity (LHT) are studied, we illustrate here the discriminating power of our set of observables using the example of the effects in models based on the MSSM. The most interesting CP asymmetries are $A_{7,8,9}$, which are not suppressed by small strong phases \cite{Bobeth:2007dw} and thus potentially of $O(1)$.

\begin{figure}[t]
\includegraphics[width=4.2cm]{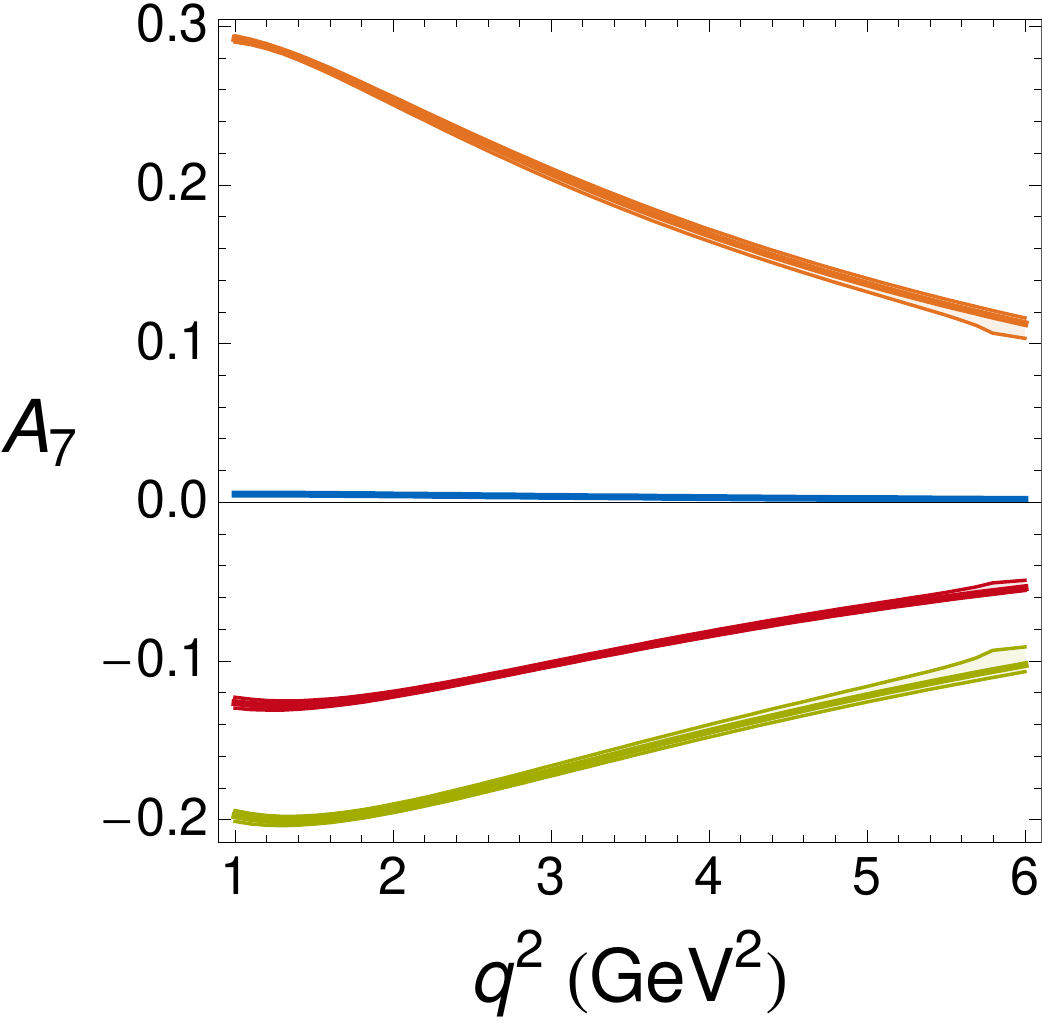}
\includegraphics[width=4.2cm]{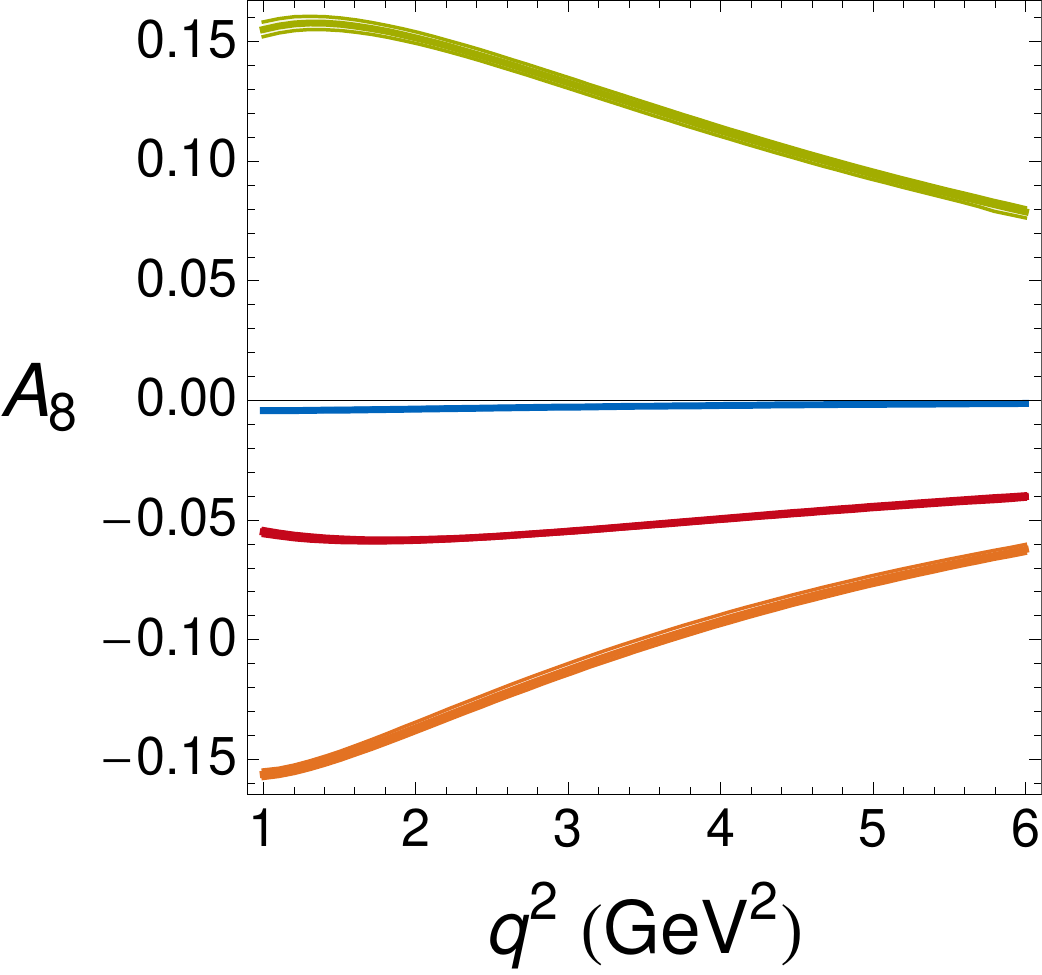}
\includegraphics[width=4.2cm]{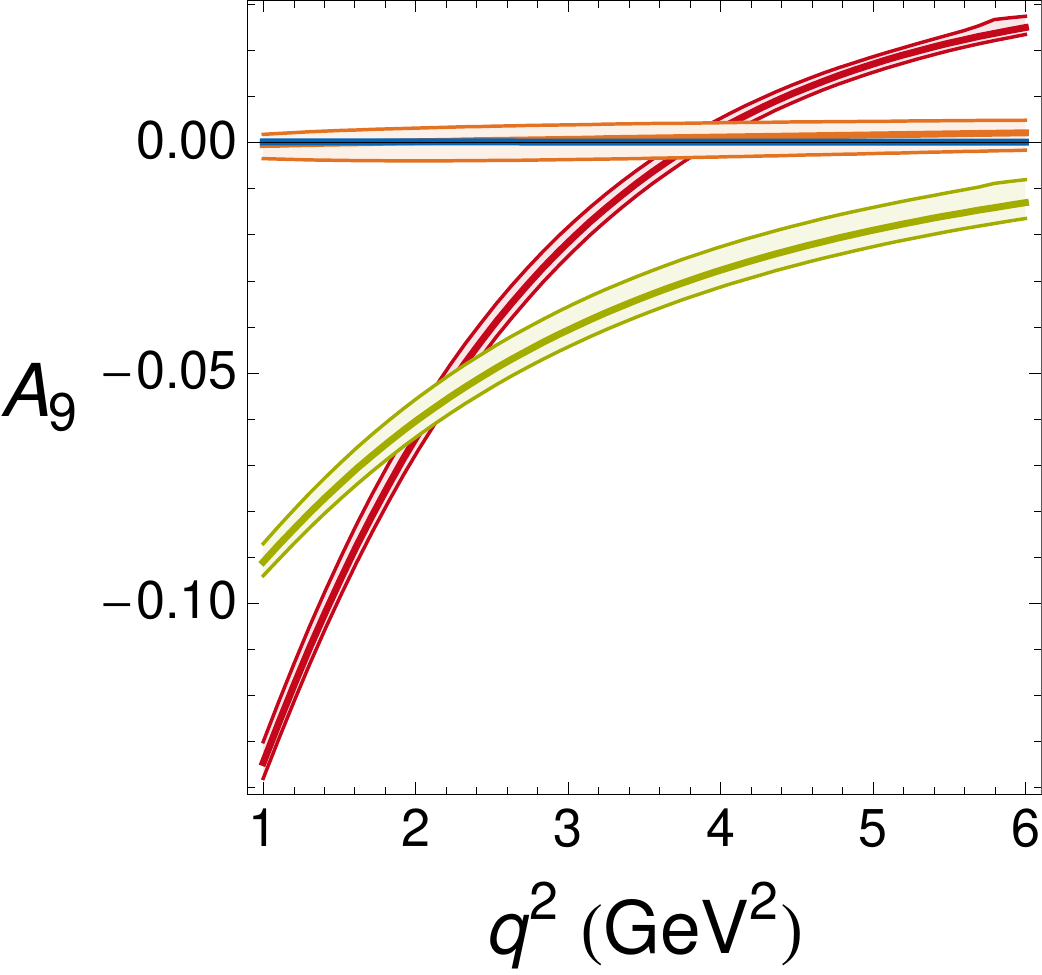}
\label{fig:FBMSSM}
\caption{Dependence of the CP asymmetries $A_{7,8,9}$ on $q^2$ in three variations of the MSSM: the Flavor Blind MSSM (orange) with $\text{Arg}(\mu A_{\tilde t})=50^\circ$, the MSSM with complex $(\delta_d)_{32}^{LR}$ mass insertion (red) and the MSSM with complex $(\delta_u)_{32}^{LR}$ mass insertion (green).  }
\end{figure}

In fig.~\ref{fig:FBMSSM} we show the effects in a Flavor Blind MSSM \cite{Altmannshofer:2008hc}, which is a modification of the Minimal Flavor Violating MSSM with additional flavor conserving CP violating phases in the soft terms. The other two scenarios are general MSSM frameworks with different mass insertions switched on. One can see that the
effects in the different observables are highly model dependent and give, combined with the other CP symmetries and asymmetries, an extraordinary tool to
discriminate between different models or parameter configurations. 

\begin{figure}[t]
\includegraphics[width=0.5 \textwidth]{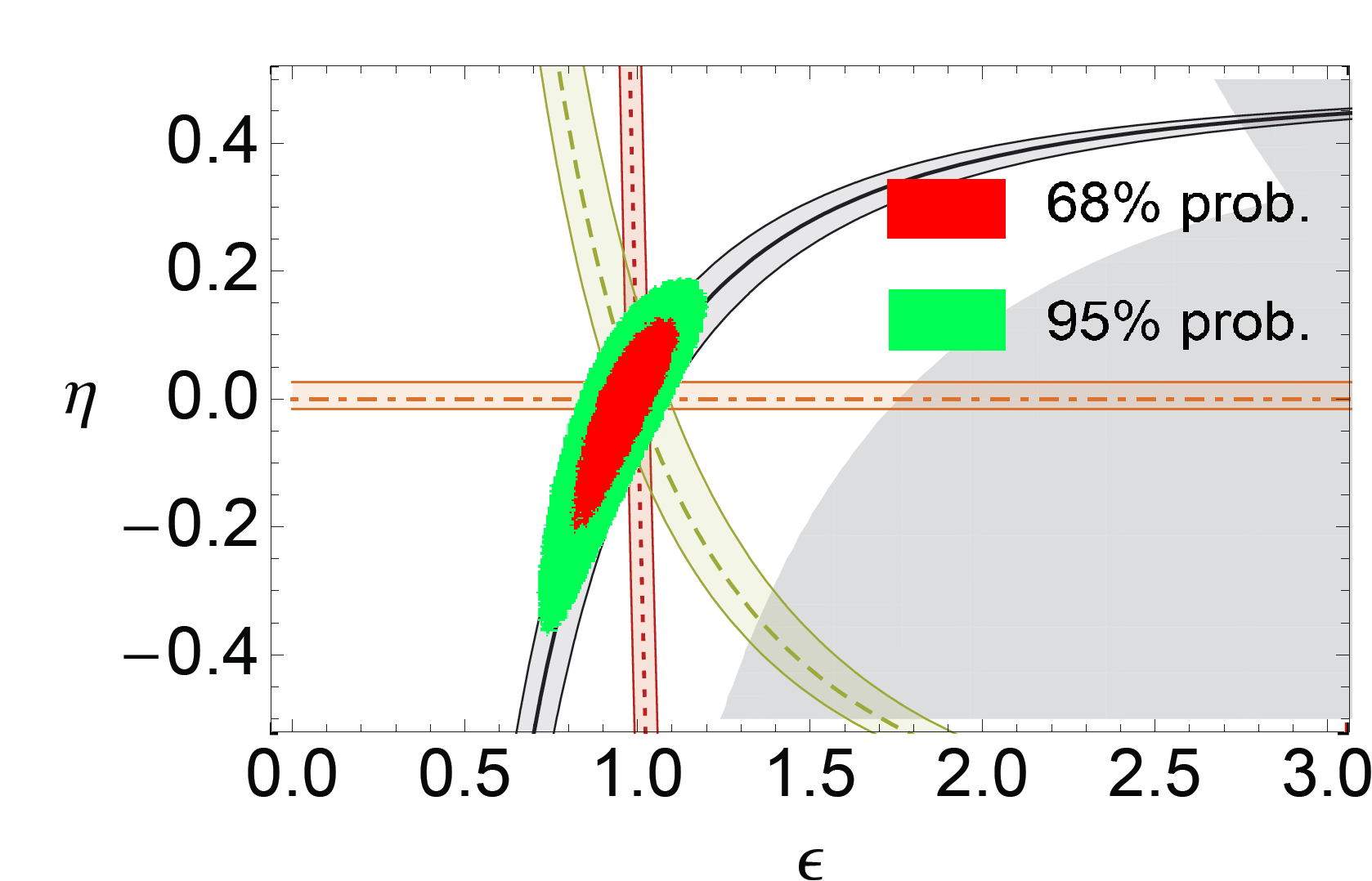}
\includegraphics[width=0.5 \textwidth]{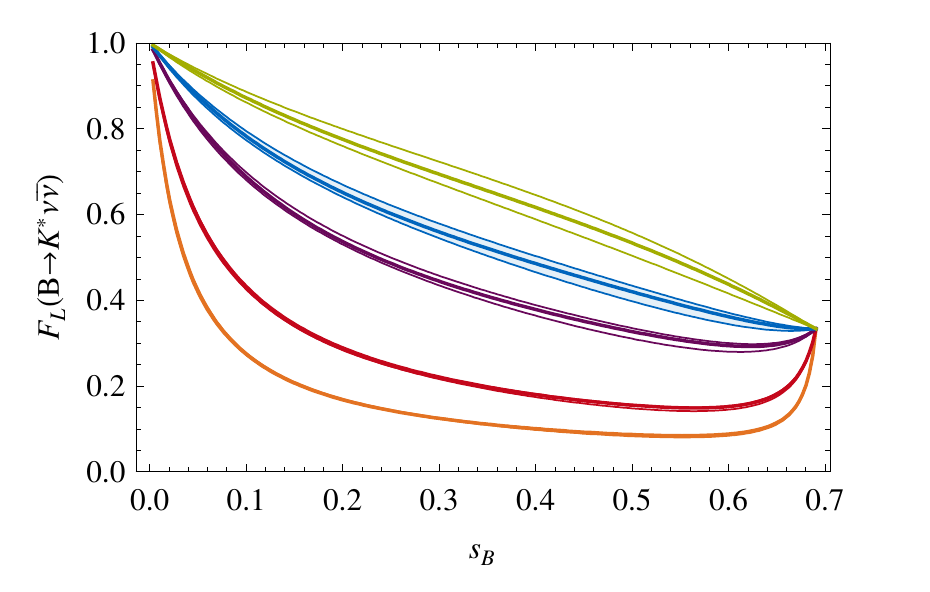}
\caption{Left: Hypothetical constraints on the $\epsilon$-$\eta$-plane, assuming all four $b \to s \nu\bar\nu$ observables have been measured with infinite precision. The error bands include the uncertainties due to the form factors in the case of the exclusive decays and the uncertainties of the CKM elements as well as the uncertainty in the 
SM Wilson coeffcient. The green band (dashed line) represents $\text{BR}(B \to K^* \nu\bar\nu)$, the black band (solid line) $\text{BR}(B \to K \nu\bar\nu)$, the red band (dotted line) $\text{BR}(B \to X_s \nu\bar\nu)$ and the orange band (dot-dashed line) $\langle F_L \rangle$. The shaded area is ruled out experimentally at the 90\% confidence level. The red and green areas are the projected sensitivity at SuperB with $75 ab^{-1}$ integrated luminosity \cite{Renga}  . Right: dependence of $F_L$ on the momentum transfer for different values of $\eta$, from top to bottom: $\eta=0.5, 0, -0.2, -0.4, -0.45$. 
}
\label{etarho}
\end{figure}
 \vspace{-0.7cm}
%%%%%%%%%%%%%%%%%%%%%%%%%%%%%%%%%%%%%%%%%%%%%%%%%%%%%%%%%%%%%%%%%%%%%%%%%%%%%%%%%%%%%%%%
\subsection{The $B\to K\nu\bar\nu$, $B\to K^*\nu\bar\nu$ and $B \to X_s\nu\bar\nu$ decays}
%%%%%%%%%%%%%%%%%%%%%%%%%%%%%%%%%%%%%%%%%%%%%%%%%%%%%%%%%%%%%%%%%%%%%%%%%%%%%%%%%%%%%%%%
%\subsubsection{effective Hamiltonian}
The effective Hamiltonian for $b\to s \nu\bar\nu$ transitions is generally given by
\begin{equation} \label{eq:Heff}
H_{\eff} = - \frac{4\,G_F}{\sqrt{2}}V_{tb}V_{ts}^*\left(C^\nu_L O^\nu_L +C^\nu_RO^\nu_R  \right) ~+~ {\rm h.c.} ~,
\end{equation}
with the operators
\begin{align}
O^\nu_{L} &=\frac{e^2}{16\pi^2}
(\bar{s}  \gamma_{\mu} P_L b)(  \bar{\nu} \gamma^\mu(1- \gamma_5) \nu)~,&
O^\nu_{R} &=\frac{e^2}{16\pi^2}(\bar{s}  \gamma_{\mu} P_R b)(  \bar{\nu} \gamma^\mu(1- \gamma_5) \nu)~.
\end{align}
%\subsubsection{quark level transition,observables}
The quark level transition $ b\to s \nu\bar\nu$ gives rise to three $B$ decays with a total of four observables. These
are the three branching ratios and one additional polarization ratio in the case of $B\to K^*\nu\bar\nu$, measuring the fraction $F_L$ of longitudinally
polarized $K^*$ mesons \cite{Altmannshofer:2009ma}. This polarization fraction can be extracted from the angular distribution in the invariant mass of the neutrino-antineutrino pair and the angle between the $K^*$ flight direction in the $B$ rest frame and the $K$ flight direction in the $K\pi$ rest frame.
%\subsubsection{reduction of hadronic/QCD hadronic errors}
A major source of uncertainties of the $ b\to s \nu\bar\nu$ based decays are the QCD/hadronic ingredients entering the calculation.
A well known problem in the inclusive decay is the $m_b$ dependence, which leads to considerable uncertainties. The traditional approach is to normalize the decay rate to the semileptonic inclusive $b \to c$ decay. On the other hand, this introduces again uncertainties through the dependence of the semileptonic phase space factor on the charm quark mass. Instead of this normalization, we use the $b$ mass evaluated in the $1S$ scheme \cite{Hoang:2000fm}, being known at a precision of $1\%$. 
For the $B\to K\nu\bar\nu$ \footnote{For a recent reconsideration of this mode see \cite{Kamenik:2009kc} and \cite{Bartsch:2009qp}.} decay we use the form factors given in \cite{Ball:2004ye}, being valid in the full physical range, while we use the already mentioned set of \cite{Altmannshofer:2008dz} for the decay $B \to K^* \nu\bar\nu$.
%\subsubsection{reduction of hadronic/QCD hadronic errors}
These improvements combined with an up to date top mass \cite{:2008vn} lead to a significantly lower prediction for $\text{BR}(B \to K^* \nu\bar\nu)$ and a considerably more accurate prediction for $\text{BR}(B \to X_s \nu\bar\nu)$, than the ones present in the literature. 

In table \ref{table1} we give a summary of our SM predictions.
\begin{table}
\centering
\begin{tabular}{|l|l|}
\hline
Observable  &  Our SM prediction \\
\hline
$\text{BR}(B \to K^* \nu\bar\nu)$ & $( 6.8^{+1.0}_{-1.1} ) \times 10^{-6}$  \\
$\text{BR}(B^+ \to K^+ \nu\bar\nu)$   & $( 4.5 \pm 0.7 ) \times 10^{-6}$ \\
$\text{BR}(B \to X_s \nu\bar\nu)$ & $( 2.7\pm0.2 ) \times 10^{-5}$ \\
$\langle F_L(B \to K^* \nu\bar\nu) \rangle$ & $0.54 \pm 0.01$  \\
\hline
\end{tabular}
\label{table1}
\caption{\small SM predictions}
\end{table}
The four observables accessible in the three different $b\to s\nu\bar\nu$ decays are dependent on the two in principle complex Wilson coefficients $C^\nu_L$ and $C^\nu_R$. However, only two real combinations of these complex quantities enter the observables \cite{Melikhov:1998ug,Altmannshofer:2009ma}:
\begin{equation}  \label{eq:epsetadef}
 \epsilon = \frac{\sqrt{ |C^\nu_L|^2 + |C^\nu_R|^2}}{|(C^\nu_L)^\text{SM}|}~, \qquad
 \eta = \frac{-\text{Re}\left(C^\nu_L C_R^{\nu *}\right)}{|C^\nu_L|^2 + |C^\nu_R|^2}~.
\end{equation}
Measurements of the four observables are then transparently represented as bands in the $\epsilon-\eta$-plane. To illustrate the theoretical cleanliness of the various observables, we show in fig. 2 the combined constraints after hypothetical measurements of the observables.
%---------------------------------------------------------------------------------------
%LHT, MSSM
%---------------------------------------------------------------------------------------
Apart from the model independent analysis, an application to specific models shows, that
NP effects in the LHT and the Randall-Sundrum model with custodial protection of left-handed Z-couplings to down type quarks are small,
as opposed to the MSSM with a generic flavor violating soft sector. Taking into account the strong
constraints from $B \to X_s \gamma$ and $B_s \to \mu^+ \mu^-$ it turns out that dominantly chargino contributions
lead to sizeable effects.
 \vspace{-0.2cm}
%%%%%%%%%%%%%%%%%%%%%%%%%%%%%%%%%%%%%%%%%%%%%%%%%%%%%%%%%%%%%%%%%%%%%%%%%%%%%%%%%%%%%%%%
\begin{theacknowledgments}
%%%%%%%%%%%%%%%%%%%%%%%%%%%%%%%%%%%%%%%%%%%%%%%%%%%%%%%%%%%%%%%%%%%%%%%%%%%%%%%%%%%%%%%%
I would like to thank W.~Altmannshofer, P.~Ball, A.Bharucha, A.~J.~Buras and D.~M.~Straub for the pleasant and fruitful collaboration. This work has been supported by the German Bundesministerium f\"ur Bildung und Forschung under contract 05HT6WOA and the Graduiertenkolleg GRK 1054 of DFG.
\end{theacknowledgments}
 \vspace{-0.2cm}
%%%%%%%%%%%%%%%%%%%%%%%%%%%%%%%%%%%%%%%%%%%%%%%%%%%%%%%%%%%%%%%%%%%%%%%%%%%%%%%%%%%%%%%%

\end{document}